\documentclass[prb,twocolumn,showpacs,preprintnumbers,superscriptaddress,amsmath,amssymb]{revtex4}

\usepackage{color,graphicx,amsmath}
\begin{document}
\title{Strong coupling between excitons in organic semiconductors and Bloch Surface Waves}

\author{Stefano Pirotta}
\author{Maddalena Patrini}
\author{Marco Liscidini}
\author{Matteo Galli}
\author{Giacomo Dacarro}
\affiliation{Dipartimento di Fisica, Universit\`{a} degli Studi di Pavia, via Bassi 6, 27100 Pavia, Italy}
\author{Giancarlo Canazza}
\affiliation{Dipartimento di Chimica e Chimica Industriale, Universit\`{a} degli Studi di Genova, via Dodecanneso 31, 16146 Genova, Italy}
\author{Giorgio Guizzetti}
\affiliation{Dipartimento di Fisica, Universit\`{a} degli Studi di Pavia, via Bassi 6, 27100 Pavia, Italy}
\author{Davide Comoretto}
\affiliation{Dipartimento di Chimica e Chimica Industriale, Universit\`{a} degli Studi di Genova, via Dodecanneso 31, 16146 Genova, Italy}
\author{Daniele Bajoni}\email[]{daniele.bajoni@unipv.it}
\affiliation{Dipartimento di Ingegneria Industriale e dell'Informazione, Universit\`{a} degli Studi di Pavia, via Ferrata 1, 27100 Pavia, Italy}

\date{\today}

\begin{abstract}
We report on the strong coupling between the Bloch surface wave supported by an inorganic multilayer structure and $J$-aggregate excitons  in an organic semiconductor. The dispersion curves of the resulting polariton modes are investigated by means of angle-resolved attenuated total reflection as well as photoluminescence experiments.  The measured Rabi splitting is 290 meV. These results are in good agreement with those obtained from our theoretical model.
\end{abstract}


\maketitle


A large number of new and compelling phenomena are based on exciton polaritons, including very low threshold lasing \cite{Imamoglu,pollaserreview}, Bose Einstein condensation \cite{Kasprzak}, and superfluidity \cite{Carusottoreview}. Polaritons are also appealing in classical and quantum nonlinear optics, since they can have strong optical nonlinearities arising from their high self-interaction potential \cite{Tignon,JacquelineWire}.

Exciton polaritons form when the interaction between excitons and the electromagnetic field is strong enough to become reversible \cite{Skolnickreview,Weisbuch} (strong coupling regime). The typical platform used to investigate this kind of polaritons consists in  one or more quantum wells embedded in a one dimensional optical microcavity \cite{Skolnickreview}. Starting from this system, one can fully confine the polariton \cite{Deveaudmesa} or control its propagation by pattering the microcavity \cite{JacquelineWire}. To this end, strong coupling has also been investigated in other planar structures, for instance exciton polaritons have been observed in GaAs slab waveguides by using dielectric \cite{PCPolPRB,L3pollaser} and metallic \cite{Whittaker} gratings \cite{Kavokin}. 

\begin{figure}[b]
\includegraphics[width= \columnwidth]{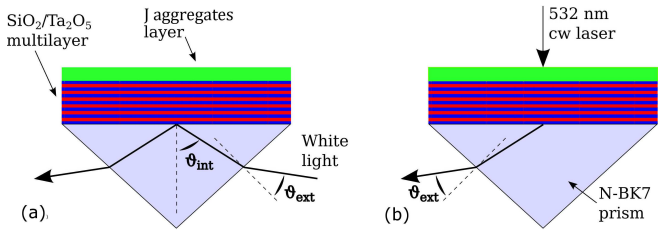}
\caption{(color online)  Scheme of the sample and of the experimental 
geometries for (a) ATR and (b) PL experiments, the drawings are not in scale. }\label{Figscheme}
\end{figure}

\begin{figure}[b]
\includegraphics[width= \columnwidth]{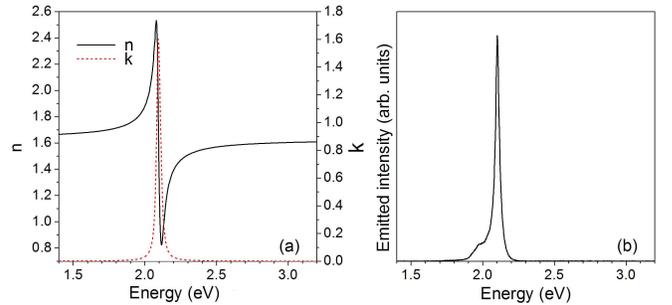}
\caption{(color online)  (a) Dispersion of the refractive 
index and extinction coefficient of the TDBC $J$-aggregates doped PVA film, measured by spectroscopic ellipsometry on a reference sample on a silicon substrate. (b) 
Photoluminescence spectrum from the film. }\label{Fig1}
\end{figure}

\begin{figure*}[t]
\includegraphics[width= \textwidth]{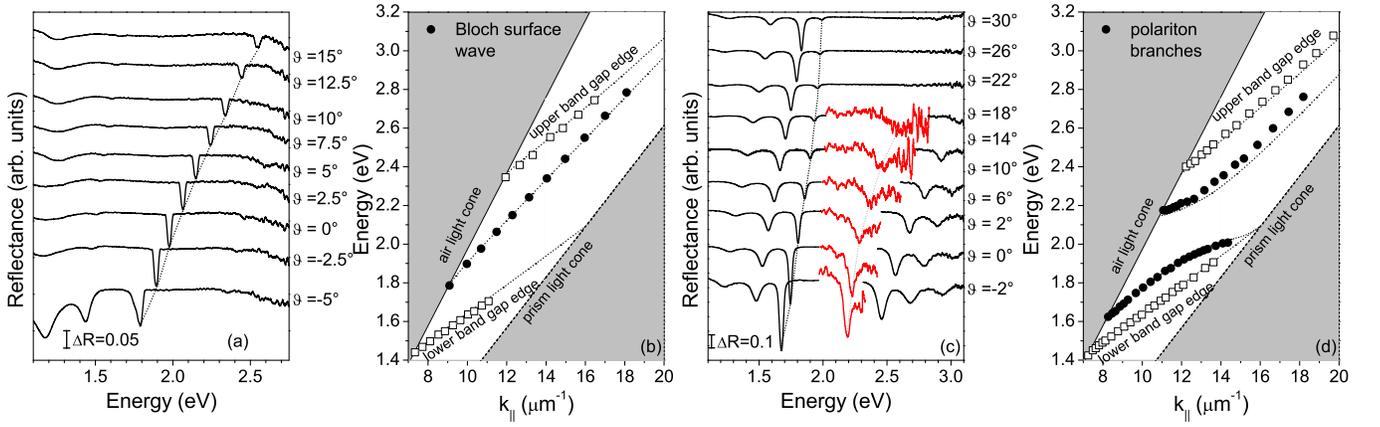}
\caption{(color online) (a) ATR spectra from the DBR structure on the prism. (b) The extracted dispersion of the BSW mode (full circle dots) and of the adjacent band gap modes (open square dots). The calculated dispersions are shown as dotted lines. (c) ATR spectra from the sample covered with the TDBC $J$ aggregates doped PVA film,  the red part of the curves is multiplied by a factor of 20 (d). The extracted dispersion of the polariton modes (full circle dots) and the adjacent band gap modes (open square dots). The calculated dispersions are shown as dotted lines. In panels (a) and (c) the spectra are vertically shifted for clarity.}\label{Fig2}
\end{figure*}

A recent theoretical proposal suggested strong coupling between excitons and Bloch Surface Waves (BSWs) \cite{BSWpol}. The latters are photonic modes that exist at the interface between a truncated photonic crystal, typically a distributed Bragg reflector (DBR), and an ideally semiinfinite dielectric medium \cite{yeh78}. BSWs are confined near the structure surface and propagate in the plane of the structure: their confinement is due to total internal reflection from the side of the homogenous dielectric and to the photonic band gap in the photonic crystal. Thus, the BSW dispersion relation lies below the light line of the top dielectric material and within the photonic band gap.  Since the electromagnetic field is highly concentrated near the sample surface \cite{liscidini09}, BSWs are appealing for enhancing the interaction between light and the  materials deposited on the sample surface \cite{liscidini09_2}, and thus they have been mainly exploited for sensing applications \cite{robertson04,descrovi07_2,pirotta13}. Yet, the small modal volume and the possibility to control BSW propagation by means of a minimal structurization of the multilayer surface \cite{BSWpol} suggest their strong potential to study the light-matter interaction in the strong coupling regime. 

In this work we report experimental evidence of strong coupling between Bloch Surface Waves supported by a Ta$_2$O$_5$/SiO$_2$ multilayer and the excitons in the $J$-aggregate form of the dye 5,5',6,6'-tetrachloro-1,1'-diethyl-3, 3'-di(4-sulfobutyl)-benzimidazolocarbocyanine (TDBC). TDBC and other dyes $J$-aggregates have been shown to undergo strong coupling in a variety of photonic platforms, from microcavities \cite{Lidzey1}, to gratings \cite{Nature,Baumberg}, metallic nanoparticles \cite{Bellessa}, opals \cite{Vardeny} and Tamm states \cite{BellessaTamm}. While not retaining some of the properties of excitons in GaAs quantum wells, they offer the great advantage of having very large oscillator strengths so that strong coupling can be easily observed at room temperature.

A sketch of the structure investigated in this work is shown in Figure \ref{Figscheme}. The sample, realized by Layertech GmbH, consists of a Ta$_2$O$_5$/SiO$_2$ multilayer grown by magnetron sputtering  on the hypotenuse of a 45$^\circ$-90$^\circ$-45$^\circ$ N-BK7 prism. It is composed of six periods of alternating Ta$_2$O$_5$ and SiO$_2$ layers (87 nm and 153 nm in thickness respectively, $\sim$0.26 refractive index contrast); the multilayer truncation consists in the two topmost layers (10 nm of Ta$_2$O$_5$ and 10nm of SiO$_2$) engineered to support a BSW in the visible spectral range. Finally, a TDBC $J$-aggregates doped PVA film was spin cast on the top surface of the multilayer, as sketched in Figure \ref{Figscheme}. TDBC (FEW chemicals, CAS number 18462-64-1) $J$-aggregates dispersed into poly(vinyl alcohol) matrix (PVA, Sigma Aldrich, M$_{\mbox{w}}$ 85000-124000, CAS number 9002-89-5) were prepared by de-ionized water solutions. 0,1 g of PVA have been added to 2 ml of a $5.2\times10^{-2}$M TDBC aqueous solution. In order to obtain a complete dissolution of PVA and an effective mixing with TDBC, the solution has been stirred and heated at 70$^\circ$C for 24 hours. 250 $\mu$l of such solution have been dynamically spin coated at 40 rps directly on the multilayer by using an SCC Novocontrol spincoater \cite{Davide1,Davide2}. The same process was also performed to coat a silicon substrate as reference sample for assessing the optical properties of the $J$-aggregates.

The optical constants of the organic semiconductor were fully characterized by considering a TDBC $J$-aggregates doped PVA film deposited on silicon substrate and by performing variable angle spectroscopic ellipsometry (VASE from J.A. Woolam Co., Inc.) in the 0.25-2.5 $\mu$m wavelength range. Ellipsometric spectra were then analyzed with the commercial software WVASE32$^{\mbox{\textregistered}}$ by assuming a three layer model: the silicon substrate, a thin silicon oxide layer, and  the TDBC $J$-aggregates doped PVA film, whose thickness and dielectric functions are the free parameters of the fit procedure. Silicon and PVA optical constants were obtained from reference samples. The best-fit value for the thickness is 42 $\pm$ 8 nm. The refractive index and extinction coefficient of the TDBC $J$-aggregates doped PVA film are shown in Figure \ref{Fig1}(a): the strong dispersive structure associated with the exciton is the sign of the high oscillator strength of the $J$-aggregate. The dispersion of the optical constants was simulated assuming an harmonic function for the dielectric response $\epsilon(\omega)=\epsilon_\infty+\omega_{LT}/(\omega-\omega_0-i\gamma)$,with $\epsilon_\infty=1.6$, $\omega_{LT}=0.4$ eV, $\omega_{0}=2.1$ eV and $\gamma=0.03$ eV.

The emission properties of the $J$-aggregates were also investigated. In Fig \ref{Fig1} (b) we show the photoluminescence (PL) spectrum of the reference TDBC/PVA layer on silicon when excited by a continuous-wave 532 nm laser source. The peak at 2.1 eV corresponds to the exciton emission and has a FWHM of $\sim$ 30 meV,  sign of the strong coherence of interaction and reduced inhomogeneous broadening of the of the TDBC $J$-aggregates doped PVA film \cite{libropolimeri}. 

\begin{figure}[t!]
\includegraphics[width= \columnwidth]{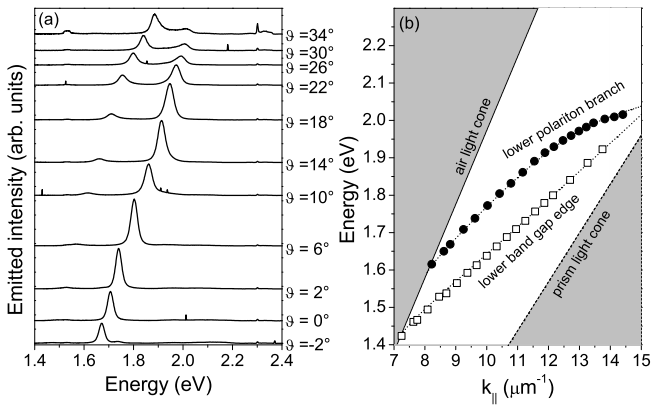}
\caption{(a) PL spectra from the prism covered with the TDBC $J$-aggregates doped PVA film, the curves are vertically shifted for clarity. (b) The extracted dispersion of the lower polariton branch (full circle dots) and the lower band gap mode (open square dots). The calculated dispersions, same as in figure \ref{Fig2} are shown as dotted lines.}\label{Fig3}
\end{figure}

The photonic modes of the system were characterized before and after the deposition of the the TDBC $J$-aggregates doped PVA film by means of angle-resolved attenuated total reflectance (ATR) in the Krestchmann configuration (see Fig. \ref{Figscheme} (a))  \cite{Galli04}. The ATR spectra for the collected transverse-electric (TE) polarized light are shown in Fig. \ref{Fig2}(a) for different external angle $\theta_{ext}$, which are determined within a resolution of $\sim$1$^{\circ}$. Several dispersive dips are clearly visible in the spectra: the sharpest one corresponds to the excitation of the BSW mode at the sample surface; the other dips are associated with bulk photonic modes, which are guided in the multilayer \cite{liscidini09_2,pirotta13}. The dispersion relations of the modes supported by the structure can be extracted from the ATR spectra by using\begin{equation}
\label{dispersion}
k_{\parallel}=\frac{E}{\hbar c}\sin\left(45^{\circ}+\frac{\theta_{ext}}{n}\right),
\end{equation} 
where  $k_{\parallel}$ is the modulus of the in-plane wave vector, $E$ is the energy, and $n$ is the prism refractive index. The results are shown in Fig. \ref{Fig2} (b) along with the theoretical curves calculated using the transfer matrix method. The BSW dispersion relation is clearly visible inside the band gap, delimited by the two band edges, and it is in excellent agreement with the theoretical curve. 

The same experiment was repeated after the deposition of the the TDBC $J$-aggregates doped PVA film. The ATR spectra and the corresponding dispersion relations are shown in Fig. \ref{Fig2} (c) and (d), respectively. In particular, the  spectra show two distinct dips, within the band gap, due to the strong coupling between the $J$-aggregate exciton and  the BSW mode. The high-energy dips are less pronounced than the low-energy dips because of the nonlinear relation between and  $k_{\parallel}$ and $\theta_{ext}$ in Eq. \ref{dispersion}, which results in the broadening of the resonance at 
high energies and large $\theta_{ext}$, and due to absorption in Ta$_2$O$_5$ above 2.4 eV. 

The dispersion relations of the different modes are plotted in Fig. \ref{Fig2} (d). While the curves corresponding to the upper and lower band edges are essentially identical to those shown in Fig. \ref{Fig2} (b),  the BSW dispersion curve obtained  without the TDBC $J$-aggregates doped PVA film  is now replaced by two branches. These show a clear anticrossing around $k_{\parallel}=$12 $\mu$m$^{-1}$, demonstrating the strong coupling regime. The Rabi splitting is around 290 meV, a value greater than those previously reported for $J$-aggregates excitons in other planar structures\cite{Lidzey1,Nature,Baumberg,BellessaTamm},  but lower than that obtained with metallic nanoparticles \cite{Bellessa}. This result is consistent with the  large field enhancement at the sample surface that can be obtained through a BSW. This is higher than what can be obtained inside a simple planar microcavity, but lower than the typical field enhancement given by plasmonic resonances \cite{BSWpol}. The dispersion relation of the polariton branches was calculated  with the same harmonic function used to fit the dielectric response of the $J$-aggregate in the ellipsometric experiments. Absorption losses were neglected for numerical reasons (i.e. $\gamma=0$). Theoretical and experimental results are in good agreement in the entire investigated energy range.

Finally, we collect photoluminescence (PL) spectra from the sample (see Fig. \ref{Fig3} (a)). The TDBC $J$-aggregates doped PVA film was pumped from the top  with a cw laser at a wavelength $\lambda_{pump}=532$ nm; PL signal was then collected through the prism in the Kretschmann configuration (see Fig. \ref{Figscheme} (b)). The wavelength $\lambda_{pump}$ is chosen to avoid spurious photoluminescence from the multilayer, which is observable when $\lambda_{pump}> 2.4$ eV.  This allows us to investigate only the emission in the region of the lower polariton branch.  While photoluminescence from the uncoupled excitons is suppressed by the photonic band gap, dispersive peaks associated with the polariton branch are clearly observable in the emission spectra. The data are in excellent agreement with the same simulated dispersion curves used to fit the dispersion relation obtained from the ATR measurements.

In conclusion we have experimentally demonstrated strong coupling between a Bloch surface wave supported by an inorganic multilayer structure and excitons in a $J$-aggregate film, thus demonstrating the existence of Bloch surface waves polaritons.  Similar results could be replicated in other systems, such as III-V semiconductors, where strong coupling is typically observed using planar microcavities. The absence of intrinsic losses of the photonic mode as well as the possibility to control the BSW propagation with a simple structurization of the sample surface point to BSWs as the ideal platform for fundamental studies and applications of exciton polaritons in integrated devices. 

This work was supported by Regione Lombardia funding through ``Fondo per la promozione di accordi istituzionali Progetti di cooperazione scientifica e tecnologica internazionale`` (Project code SAL-11), by CNISM funding through the INNESCO project ``PcPol", by MIUR funding through the FIRB ``Futuro in Ricerca" project RBFR08XMVY, from the foundation Alma Mater Ticinensis. Work in Genova is partially supported by Progetto di Ricerca di Ateneo 2012 and by the Ministry of the University and Scientific and Technological Research through the PRIN 2010-2011 project (2010XLLNM3).

\end{document}